\documentclass[sigconf,table]{acmart}

\AtBeginDocument{%
  \providecommand\BibTeX{{%
    \normalfont B\kern-0.5em{\scshape i\kern-0.25em b}\kern-0.8em\TeX}}}

\setcopyright{acmcopyright}
\copyrightyear{2022}
\acmYear{2022}

\acmConference[MSR 2022]{MSR '22: Proceedings of the 19th International Conference on Mining Software Repositories}{May 23?24, 2022}{Pittsburgh, PA, USA}

\usepackage{algorithmic}
\usepackage{graphicx}
\usepackage{textcomp}
\usepackage{xcolor, colortbl}
\usepackage{booktabs}
\usepackage{multirow}
\usepackage{xspace}
\usepackage{caption}
\usepackage{subcaption}
\usepackage{footnote}
\usepackage{url}
\usepackage{tcolorbox}
\usepackage{booktabs}
\usepackage{tikz}
\usepackage{fontawesome}
\usepackage{adjustbox}

\usepackage{listings}

\definecolor{Gray}{gray}{0.9}

\def\BibTeX{{\rm B\kern-.05em{\sc i\kern-.025em b}\kern-.08em
    T\kern-.1667em\lower.7ex\hbox{E}\kern-.125emX}}

\newcommand{\ie}{\emph{i.e.,}\xspace}
\newcommand{\eg}{\emph{e.g.,}\xspace}
\newcommand{\etc}{etc.\xspace}
\newcommand{\etal}{\emph{et~al.}\xspace}
\newcommand{\secref}[1]{Section~\ref{#1}\xspace}
\newcommand{\figref}[1]{Fig.~\ref{#1}\xspace}

\newcommand{\tabref}[1]{Table~\ref{#1}\xspace}

\usepackage[T1]{fontenc}

\usepackage{color}

\definecolor{pblue}{rgb}{0.13,0.13,1}
\definecolor{pgreen}{rgb}{0,0.5,0}
\definecolor{pred}{rgb}{0.9,0,0}
\definecolor{pgrey}{rgb}{0.46,0.45,0.48}
\definecolor{gray05}{gray}{0.95}

\usepackage{listings}
\newboolean{showcomments}
\setboolean{showcomments}{true}

\ifthenelse{\boolean{showcomments}}
  {\newcommand{\nb}[2]{
    \fbox{\bfseries\sffamily\scriptsize#1}
    {\sf\small$\blacktriangleright$\textit{#2}$\blacktriangleleft$}
   }
  }
  {\newcommand{\nb}[2]{}
  }

\title{To What Extent do Deep Learning-based Code Recommenders Generate Predictions by Cloning Code from the Training Set?}

\author{Matteo Ciniselli}
\affiliation{%
  \institution{\small{SEART @ Software Institute\\Universit\`a della Svizzera italiana}}
 \country{Switzerland}
}
\author{Luca Pascarella}
\affiliation{%
  \institution{\small{SEART @ Software Institute\\Universit\`a della Svizzera italiana}}
  \country{Switzerland}
}
\author{Gabriele Bavota}
\affiliation{%
  \institution{\small{SEART @ Software Institute\\Universit\`a della Svizzera italiana}}
  \country{Switzerland}
}



\copyrightyear{2022} 
\acmYear{2022} 
\setcopyright{acmcopyright}\acmConference[MSR '22]{19th International Conference on Mining Software Repositories}{May 23--24, 2022}{Pittsburgh, PA, USA}
\acmBooktitle{19th International Conference on Mining Software Repositories (MSR '22), May 23--24, 2022, Pittsburgh, PA, USA}
\acmPrice{15.00}
\acmDOI{10.1145/3524842.3527938}
\acmISBN{978-1-4503-9303-4/22/05}

\begin{document}

\begin{abstract}
Deep Learning (DL) models have been widely used to support code completion. These models, once properly trained, can take as input an incomplete code component (\eg an incomplete function) and predict the missing tokens to finalize it. GitHub Copilot is an example of code recommender built by training a DL model on millions of open source repositories: The source code of these repositories acts as training data, allowing the model to learn ``how to program''. The usage of such a code is usually regulated by Free and Open Source Software (FOSS) licenses, that establish under which conditions the licensed code can be redistributed or modified. As of Today, it is unclear whether the code generated by DL models trained on open source code should be considered as ``new'' or as ``derivative'' work, with possible implications on license infringements. In this work, we run a large-scale study investigating the extent to which DL models tend to clone code from their training set when recommending code completions. Such an exploratory study can help in assessing the magnitude of the potential licensing issues mentioned before: If these models tend to generate new code that is unseen in the training set, then licensing issues are unlikely to occur. Otherwise, a revision of these licenses urges to regulate how the code generated by these models should be treated when used, for example, in a commercial setting. Highlights from our results show that $\sim$10\% to $\sim$0.1\% of the predictions generated by a state-of-the-art DL-based code completion tool are Type-1 clones of instances in the training set, depending on the size of the predicted code. Long predictions are unlikely to be cloned.
\end{abstract}

\maketitle

\eject

\section{Introduction} \label{sec:intro}

Code completion is nowadays a ``killer'' feature for Integrated Development Environments (IDEs)~\cite{Bruch:fse2009,Robb2010a,kim2020code}, and it can substantially speed up implementation activities by recommending to the developer the next code token(s) to write~\cite{han2009code,han2011code}. Early code completion techniques were merely based on alphabetically sorted lists of the next token to write given the characters already typed. 

Over the years, researchers pushed the capabilities of these tools by mainly focusing on their prediction accuracy (\ie their ability to predict the next tokens to write correctly). This has been achieved, for example, by (i) exploiting contextual information such as the code surrounding the statements on which code completion is triggered~\cite{Bruch:fse2009,Robb2010a} and/or or the recent code change history~\cite{Robb2010a}; and (ii) moving towards more advanced techniques such as Deep Learning (DL) models, that have been widely applied to recommend code tokens~\cite{Karampatsis:DLareBest,kim2020code,alon2019structural,svyatkovskiy2020intellicode,White2015}.

Most of the proposed tools were limited to the prediction of just a single token, and only recent studies focused on predicting multiple contiguous tokens or even entire statements~\cite{alon2019structural,svyatkovskiy2020intellicode,ciniselli2021empirical}. Building on top of this research, OpenAI and GitHub presented GitHub Copilot~\cite{chen2021evaluating}, a code assistant that achieved state-of-the-art performance in code recommendation. Being trained on more than 150Gb of data from over 50M public GitHub repositories, Copilot is able to even recommend entire methods just starting from their signature. 

The rise of DL to support code-related tasks was possible thanks to the unprecedented amount of development data being publicly available (\eg from GitHub) and that can be used for training the DL models. The vast majority of this data comes from open source repositories and hence its usage is regulated by specific Free and Open Source Software (FOSS) licenses (\eg Apache 2.0\footnote{\url{https://www.apache.org/licenses/LICENSE-2.0}}, GPL v3.0\footnote{\url{https://www.gnu.org/licenses/gpl-3.0.html}}). These licenses regulate how the licensed source code can be redistributed or modified to create \emph{derivative work}. However, they were not meant to define whether (i) the licensed code can be used for training DL-based code recommender, nor (ii) the code recommended by the trained models should be considered as \emph{derivative work} or, instead, as new code ``written from scratch''. Such a discussion is part of a non-trivial debate that will likely result in the update of FOSS licenses in the near future.
 
In this paper, we are interested in investigating a related but more concrete research question (RQ): 
\begin{quote}
\emph{To what extent do DL-based code recommender systems generate predictions by cloning code snippets from the training set?} 
\end{quote}

The relevance of this RQ is dictated by the following observation: If DL-based code recommenders tend to clone code snippets from the training set when generating predictions, it would be difficult to consider the recommendations as new and original code (rather than as derivative work). Although it is reasonable that the generated recommendations are influenced by the training dataset (\eg by often reusing frequent code identifiers), the problem would be more relevant if these techniques perform a verbatim copy of the training code, possibly spanning several statements.


To answer our research question, we need: (i) a DL-based code completion technique able to recommend several tokens/statements; (ii) access to the dataset(s) used to train the DL model; and (iii) access to the predictions generated by the technique on a test set. Unfortunately, such an analysis is not possible using GitHub Copilot since the training dataset is not publicly available. For this reason, we adopt the Text-To-Text-Transfer-Transformer (T5), a DL model that has been used to support several code-related tasks~\cite{Mastropaolo:icse2021}, including code completion~\cite{ciniselli2021empirical}. 

The T5 model exploits two training phases, namely pre-training and fine-tuning~\cite{raffel2019exploring}, thus requiring two training datasets. In the pre-training the model acquires knowledge about a language of interest (\eg Java code) through a self-supervised task, in which the model is asked to guess randomly masked tokens in an input instance (\eg a Java method). In the fine-tuning phase, the model is then specialized for the specific task of interest (\eg bug-fixing, predicting entire code statements).

We start from the pre-trained T5 made available by Ciniselli \etal~\cite{ciniselli2021empirical}, who also made publicly available the pre-training dataset. The model has been pre-trained on $\sim$2M Java methods. Then, we fine-tuned this model on a new dataset we built featuring 841,318 Java methods to train the model in predicting non-trivial code snippets composed by several statements. Indeed, predictions only spanning a few code tokens are not interesting for our goal, since they cannot really be defined as ``clones'' even if copied from the training dataset. We built four versions of the fine-tuning dataset, each one applying a different strategy to remove near duplicates~\cite{allamanis2019adverse}, including more and less strict strategies. Indeed, one of the assumptions we want to test is that a higher presence of near-duplicates in the training set pushes the model to copy more, since it observes over and over the same pattern.

Once trained the four models on the different datasets, we used them to generate predictions on a test set of $\sim$16k instances (\ie $\sim$16k Java methods with missing code statements that must be guessed by the model). Then, we used the \textsc{Simian} clone detector~\cite{simian} to identify Type-1 (\ie exact copies) and Type-2 (\ie copied code with changes to identifiers and types) clones between the generated predictions and both the pre-training and fine-tuning training datasets. Also, we checked whether the model tends to ``copy'' more when (i) dealing with predictions having different length, since we expect smaller predictions to more likely result in clones; and (ii) generating correct or wrong predictions.

The achieved results show that $\sim$10\% of the generated predictions represent a Type-1 clone of code in the training data, with this percentage going up to $\sim$80\% when considering Type-2 clones. However, the percentage of clones steadily drops when the model is asked to predict several statements, with basically no clones generated when the prediction comprises at least four lines of code. 

We also found that correct predictions are more likely to contain clones of code present in the training datasets. Such a result stresses once more the importance of a proper dataset cleaning to avoid overlaps between the training and the test datasets.

\section{Study Design} \label{sec:design}

\definecolor{gray50}{gray}{.5}
\definecolor{gray40}{gray}{.6}
\definecolor{gray30}{gray}{.7}
\definecolor{gray20}{gray}{.8}
\definecolor{gray10}{gray}{.9}
\definecolor{gray05}{gray}{.95}

\newlength\Linewidth
\def\findlength{\setlength\Linewidth\linewidth
	\addtolength\Linewidth{-4\fboxrule}
	\addtolength\Linewidth{-3\fboxsep}
}
\newenvironment{rqbox}{\par\begingroup
	\setlength{\fboxsep}{5pt}\findlength
	\setbox0=\vbox\bgroup\noindent
	\hsize=0.95\linewidth
	\begin{minipage}{0.95\linewidth}\normalsize}
	{\end{minipage}\egroup
	\textcolor{gray20}{\fboxsep1.5pt\fbox{\fboxsep5pt\colorbox{gray05}{\normalcolor\box0}}}
	\endgroup\par\noindent
	\normalcolor\ignorespacesafterend}

The \emph{goal} of this study is to understand the extent to which DL-based code recommenders are prone to suggest code snippets being \emph{clones} of instances present in the training set. The \emph{context} is represented by the four training datasets described in Section~\ref{sec:dataset-construction} and by the snippets of code generated by a state-of-the-art code recommender built on top of the Text-To-Text-Transfer-Transformer (T5) \cite{ciniselli2021empirical}.

Our study aims at answering the following research question:

\begin{center}	
	\begin{rqbox}
		\textbf{RQ.} \textit{To what extent do DL-based code recommender systems generate predictions by cloning code snippets from the training set?}
	\end{rqbox}	 
\end{center}

Our RQ sheds some light on the implications of using recommendations generated by DL-based code recommenders in terms of possible licensing issues occurring when using the generated recommendations in the development of commercial software. We answer our RQ by training a DL-based code recommender to then use it in thousands of code completion tasks (\ie an incomplete code is provided as input to the model, that is required to generate the missing code). The generated recommendations are then compared to the instances in the training set looking for clones.


\subsection{Study Context: DL-based Code Recommender}\label{sec:code-recommender}

The first step in the selection of the study context is the code recommender to use. The latest and likely more effective code recommender available, namely GitHub Copilot \cite{chen2021evaluating}, was not an option for our study. Indeed, while we have access to Copilot, the dataset used to train it is not publicly available, making it impossible to check whether the code it recommends is a clone of the instances in its training set. For this reason, we focused on another DL-based code recommender recently proposed by Ciniselli \etal \cite{ciniselli2021empirical}. 

The authors trained a T5 model supporting code completion in Java at method-level granularity. This means that once trained the model can take as input an \emph{incomplete} Java method (\ie a method missing one or more contiguous code tokens) and it can predict the missing tokens. The T5 has been trained in two phases. The first, named pre-training, aims at providing the model with a general knowledge about the language of interest. Ciniselli \etal \cite{ciniselli2021empirical} used a classic denoising pre-training task in which the model has been fed with $\sim$2M Java methods having 15\% of their tokens (even non-contiguous ones) randomly masked, with the model asked to predict them. Once pre-trained, the model has been fine-tuned to support code completion at different levels: (i) \emph{token-level}, in which the last $x$ tokens (with 1 $\leq$ $x$ $\leq$ 10) of a given statement have been masked and must be predicted; (ii) \emph{construct-level}, in which the masking has been focused on specific code constructs, such as the conditions of {\tt if} statement or of a {\tt while}/{\tt for} loop, the parameters of method calls, \etc (see \cite{ciniselli2021empirical} for details); and (iii) \emph{block-level}, in which entire code blocks\footnote{A code block is defined as the code enclosed between two curly brackets.} up to two statements are masked. 

The fine-tuning has been performed on different datasets, for a maximum of 750k instances (methods) involved. 

In the reported evaluation, the T5 achieved correct predictions ranging from $\sim$29\%, obtained when asking the model to guess entire blocks, up to $\sim$69\%, reached in the simpler scenario of few tokens masked from the same code statement \cite{ciniselli2021empirical}.

We start from the pre-trained model made available by Ciniselli \etal in their replication package~\cite{TSE2021replication}, but we fine-tune it from scratch on a new dataset described in \secref{sec:dataset-construction}. Indeed, the fine-tuned model used by Ciniselli \etal \cite{ciniselli2021empirical} was not appropriate for our study, since they trained the T5 to predict at most two code statements. In our work, we are interested in identifying code clones in the generated predictions and, for this reason, we want to train the T5 to generate also longer prediction. Indeed, clones represented by a few code tokens or a single statement are unlikely to be relevant, but mostly due to the syntactic sugar of programming languages.

\subsection{Study Context: Datasets Construction} \label{sec:dataset-construction}

To create our fine-tuning dataset we started from the \textsc{CodeSearchNet} dataset provided by Husain \etal~\cite{Java:CodeSearchNet}. \textsc{CodeSearchNet} features over $\sim$1.5M Java methods collected from open source non-forked GitHub repositories. On top of them, we added the $\sim$2.2M Java methods made available by Ciniselli \etal~\cite{ciniselli2021empirical} and collected from GitHub Android repositories. From the overall set, we removed methods (i) having less than three lines of code, since they do not allow to train the model on ``code completion scenarios'' that are interesting in the context of code clones (\ie when asking the model to generate several missing code statements); (ii) having the name containing the \emph{test} substring in an attempt to remove test methods; and (iii) containing non-ASCII characters (\eg Chinese text) to avoid confusing the model during training. We then excluded all methods having a duplicate in the pre-training dataset used by Ciniselli \etal, obtaining the final set of 841,318 instances. 



\begin{table}[t]
	\small
	\centering
	\caption{Threshold (Th) configurations and number of methods after the filtering.\vspace{-0.2cm}}
	\label{tab:threshold}
	\begin{tabular}{lrrr}
	\toprule
	{\bf Name} & {\bf Set Th.} & {\bf Multiset Th.} & {\bf \# of Methods}\\\midrule
	Very Weak & 0.8 & 0.7 & 838k\\
	Weak & 0.6 & 0.5 & 711k\\
	Strong &0.4 & 0.3 & 483k\\
	Very Strong & 0.3 & 0.2 & 273k\\
\bottomrule
\end{tabular}
\vspace{-0.3cm} 
\end{table}

As a final step, we applied the approach defined by  Allamanis~\cite{allamanis2019adverse} and Lopes \etal~\cite{lopes2017dejavu} and used in the construction of the \textsc{CodeSearchNet} database~\cite{Java:CodeSearchNet} to identify near-duplicates in our fine-tuning dataset. The idea behind this approach is to look at the percentage of overlapped tokens between two instances (\ie methods) and identify them as near-duplicates if such an overlap exceeds a specific threshold. The code implementing such an approach is publicly available \cite{Deduplication}, and allows setting two thresholds making the identification of near-duplicates weaker (\ie only very similar methods are considered clones) or stronger. Both thresholds define the percentage of shared tokens needed to consider two methods as duplicates. However, the first threshold (\emph{Multiset Threshold}) considers the entire list of tokens in a method (including duplicates), while the second (\emph{Set Threshold}) is computed on the unique set of tokens in the compared methods.

We experiment with the four different combination of thresholds reported in \tabref{tab:threshold}, that resulted in four different fine-tuning datasets composed by the \emph{\# of Methods} instances not considered near-duplicates (out of the starting $\sim$850k). In the following we refer to the four combinations as \emph{very weak} (\ie the combination of thresholds rarely reports methods as near-duplicates), \emph{weak}, \emph{strong}, and \emph{very strong} (\ie the combination of thresholds frequently reports methods as near-duplicates) and to the resulting datasets as \emph{dataset$_{very\_weak}$}, \emph{dataset$_{weak}$}, \emph{dataset$_{strong}$}, and \emph{dataset$_{very\_strong}$}. The choice of experimenting with four different combinations was dictated by the following assumption: It is possible that a training dataset featuring several near-duplicates (such as \emph{dataset$_{very\_weak}$}) pushes the model to ``copy'' more from the training set as compared to a dataset featuring almost no near-duplicates (\emph{dataset$_{very\_strong}$}). Indeed, seeing a code snippet over and over across several training sample may push the model to use that coding pattern more when generating the predictions. This is just an assumption we validate in our study.

As described in \secref{sec:model-training}, the four datasets have been used to fine-tune four T5 models starting from the pre-trained model by Ciniselli \etal~\cite{ciniselli2021empirical}.

\subsection{Model Training} \label{sec:model-training}

We do not discuss the architectural details of T5, which have been widely documented by Raffel \etal~\cite{raffel2019exploring} to better focus on implementation choices that are relevant for this study. The specific architecture we use is the T5$_{small}$ presented in \cite{raffel2019exploring}.

\textbf{Tokenization.} T5 uses a tokenizer to pre-process the input and generate the stream of output tokens. As it will be clear later, in our study the model will be fed with a Java method in which specific code statements have been masked (input) and will be asked to generate the masked statements (output). The adopted tokenization strategy impacts the vocabulary that the model can exploit to generate the output. Previous work studied the pros and cons of different tokenization strategies, spacing from word-~\cite{domingo2018much} to character-level~\cite{mcnamee2004character} tokenization. The former is prone to the out-of-vocabulary problem~\cite{karampatsis2020big} while the latter struggles with long inputs. To mitigate these limitations, Sennrich \etal~\cite{sennrich2015neural} introduced the Byte Pair Encoding (BPE) tokenization that splits the input into a sequence of subtokens (\eg it can split the word ``\emph{string}'' into two parts ``\emph{str}'' and ``\emph{ing}'', where ``\emph{str}'' is the root and ``\emph{ing}'' is the specialization). Through this sub-splitting, BPE has the advantage of reducing the vocabulary size (as well as the out-of-vocabulary problem) and allows the model to generate composed words. For example, when applied to source code the model can generate unseen identifiers whose name is a composition of the sub-tokens present in the vocabulary. The T5 exploits a \emph{SentencePiece} \cite{DBLP:journals/corr/abs-1808-06226} tokenizer which uses the same idea behind BPE. Ciniselli \etal~\cite{ciniselli2021empirical} trained a 32k-word SentencePiece tokenizer on the pre-training dataset. Since we reuse their pre-training dataset, we also reused their trained \emph{SentencePiece} tokenizer.



\begin{table}[t]
	\centering
	\caption{Study datasets. We reported the number of instances for the training and the number of distinct methods.}
	\label{tab:dataset}
	\begin{tabular}{llrr}
	\toprule
	{\bf Filtering}  & \multirow{2}{*}{\bf Dataset} & \multirow{2}{*}{\bf \#Instances} & {\bf \#Distinct} \\
	{\bf Level} & & & {\bf Methods} \\
	\midrule
	          	          
	Very Weak 	&  \multirow{4}{*}{Training}     & 1.2M   & 797k \\
                      \addlinespace[0.08cm]
	Weak 		&      & 1.02M & 671k  \\
	                \addlinespace[0.08cm]
	Strong 		&      & 703k   & 446k \\
	                    \addlinespace[0.08cm]
	Very Strong &      & 390k   & 240k    \\\midrule
	    \multirow{2}{*}{All levels}                               & Evaluation & 15,783 & 8,560 \\
	    				 & Test 	      & 15,742 & 8,506 \\
\bottomrule
\end{tabular}
\end{table}

\textbf{Datasets masking and splitting.} To train T5 with the goal of generating non-trivial snippets of code composed by multiple statements, we adapted one of the \emph{block-level} masking scenario proposed by Ciniselli \etal~\cite{ciniselli2021empirical}. 

In particular, we apply the following processing to each instance (\ie Java method) in each of the four datasets (\ie \emph{dataset$_{very\_weak}$}, \emph{dataset$_{weak}$}, \emph{dataset$_{strong}$}, and \emph{dataset$_{very\_strong}$}). In each method, we identify all blocks (\ie code snippets enclosed between two curly brackets) composed by at least two and at most six statements. Given a method including $n$ of such blocks, we generate $n$ versions of it, each having one of the $n$ blocks masked with the special token \textsc{<extra\_id\_}\oldstylenums{0}\textsc{>}. During training and testing the model is then fed with methods including a masked block and it is required to generate it. The lower and upper bound for the block size (\ie two and six lines) are defined, based on our experience, to obtain code generation tasks that are non-trivial (at least two statements), relevant in the context of cloning, and addressable given the employed DL model, training dataset, and hardware resources (at most six statements). To account for trivial statements, we count only statements composed by at least two characters to exclude, for example, statements including only a closing parenthesis. Thus, a block composed by one 10-character statement and one 1-character statement is not considered in our study, since it does not match the ``two statements'' lower bound.

Starting from these instances, we created the training, evaluation, and test datasets in \tabref{tab:dataset}. First, it is important to clarify that the splitting across training, evaluation, and test sets was not random to avoid biasing the learning. To explain this point, let us go back to the way in which we created the instances (\ie Java methods with a masked code block) for our dataset. Given a method $m$ having $n$ blocks composed by two to six statements, we added in the dataset $n$ versions of $m$, each having one and only one block masked. Suppose that $m$ contains two blocks $b_1$ and $b_2$, thus leading to two versions of $m$: one in which $b_1$ is masked ($m_{b_1}$) and $b_2$ is not and \emph{vice versa} ($m_{b_2}$). With a random splitting, it could happen that $m_{b_1}$ is assigned to the training set and $m_{b_2}$ to the test set. However, in $m_{b_1}$ the $b_2$ is not masked. Thus, when the model has to guess the tokens masked in $m_{b_2}$ it would have the solution in the training set, resulting in boosted prediction performance. For this reason, the splitting was done at ``method-level'' rather than ``instance-level''. This means that all instances related to a method $m$ can belong to only one of the three sets (training, evaluation, testing). The number of unique methods in each set is reported in \tabref{tab:dataset} in the \emph{\# Distinct Methods} column, while the number of instances these methods generated is represented in the \emph{\# Instances} column. 

Second, as previously said, in our study we want to investigate whether the prevalence of near-duplicates in the training dataset (\ie the different filtering levels we applied, from \emph{very weak} to \emph{very strong}) plays a role in the generation of code clones. For this reason, while the training sets for the four datasets can be different, with \emph{dataset$_{very\_weak}$} being the largest one (\ie containing more near-duplicates), we wanted to have the same evaluation and test sets for all datasets, to allow for a fair comparison. For this reason, we built the evaluation and the test dataset with $\sim$15k instances each. Those instances meet the following requirements: (i) they have been extracted from the methods that were present in all four datasets, independently from the near-duplicates filtering level that has been applied; and (ii) they are balanced in terms of prediction difficulty, featuring $\sim$3k instances in which a two-statement block is masked, $\sim$3k in which a three-statement block is masked, \etc, up to six-statement blocks masking.



\begin{table}[t]
	\centering
	\caption{Hyperparameters Tuned for the T5 Models.}
	\label{tab:hyperparameters}
	\begin{tabular}{ll}
		\toprule
		\textbf{Learning Rate Type} & \textbf{Parameters}                             \\
		\midrule
		Constant (C-LR)             & $\mathit{LR} = 0.001$                           \\
		Slanted Triangular (ST-LR)  
		                            & $\mathit{LR}_{\mathit{starting}} = 0.001$       \\
		                            & $\mathit{LR_{\mathit{max}}} = 0.01$             \\
		                            & $\mathit{Ratio} = 32$                           \\
		                            & $\mathit{Cut} = 0.1$                            \\
		Inverse Square Root (ISQ-LR) 
		                            & $\mathit{LR}_{\mathit{starting}} = 0.01$        \\
		                            & $\mathit{Warmup} = 10,000$                      \\
		Polynomial Decay (PD-LR)    
		                            & $\mathit{LR}_{\mathit{starting}} = 0.01$        \\
		                            & $\mathit{LR}_{\mathit{end}} = 1\mathrm{e}{-06}$ \\
		                            & $\mathit{Power} = 0.5$                          \\
		\bottomrule
	\end{tabular}
\end{table}

\textbf{Hyperparameters tuning.} We started from the pre-trained T5 model by Ciniselli \etal~\cite{ciniselli2021empirical}. Then, to find the best T5 configuration for the fine-tuning, we followed the hyperparameters tuning procedure previously used by Mastropaolo \etal~\cite{Mastropaolo:icse2021}. In particular, we trained four different configurations of the T5 that differ for the type of learning rate applied during the training (see \tabref{tab:hyperparameters} for the four configurations). Once trained, each model has been run on the evaluation set to assess its performance in terms of percentage of correct predictions, namely the percentage of instances in the evaluation set for which the model managed to exactly predict the masked code. This process has been performed on the largest dataset (\ie  \emph{dataset$_{very\_weak}$}), assuming that the best configuration identified on it represents the best choice also for the other three datasets. Each model has been fine-tuned for 50k (corresponding to $\sim$10 epochs given the used batch size of 256). The \emph{Slanted Triangular} learning rate obtained the best results and, thus, has been used in the study to train our models.



\begin{table}[t]
	\centering
	\caption{Number of finetuning steps and best checkpoint found.}
	\label{tab:finetuning}
	\begin{tabular}{lrr}
	\toprule
	{\bf Filtering}  & {\bf \# Training} & {\bf Best}\\
	{\bf Level} & {\bf Steps} &{\bf Checkpoint}\\
	\midrule
	Very Strong &152k&20k\\
	Strong &275k&40k\\
	Weak &400k&225k\\
	Very Weak &470k&450k\\
	                    
\bottomrule
\end{tabular}
\end{table}

\textbf{Fine tuning.} We fine-tuned T5 on each of the four datasets for $\sim$100 epochs, by varying the number of fine-tuning steps based on the dataset size (\ie larger datasets require more steps to reach 100 epochs, see \emph{\# Training Steps} in \tabref{tab:finetuning}). To avoid overfitting, we saved the trained models every 5k steps and identified the best one (see \emph{Best Checkpoint} in \tabref{tab:finetuning}) on the evaluation set in terms of percentage of correct predictions. 

These are the models we will use to generate the predictions on the test set and analyze the overlap in terms of code clones between the generated predictions and the code in the training sets.

\subsection{Data Collection And Analysis}\label{sec:data-collection}

Once trained the four models (one on each training dataset), we run them on the test set that, as previously explained, is the same for all four models. This results in the models generating the code blocks predicted as needed to fill the masked code. These blocks of generated code are the ones we want to contrast against the training set used for each model to identify code clones. 

To identify the clone detector to use, we started from the literature review by Rattan \etal~\cite{rattan2013software}, looking at the list of tools documented by the authors. Several of them are no longer available (\eg \cite{higo2009enhancing}, \cite{GveroKKP13}), while others do not work on syntactically incorrect code that may be generated by the T5 models (\eg \cite{roy2008nicad}, \cite{CPD}). Given the study's constraints, we used the \textsc{Simian} clone detector~\cite{simian}. Besides being very robust and scalable, \textsc{Simian} can identify Type-1 and Type-2 clones \cite{roy2007survey}. \textsc{Simian} works at line-level granularity, looking for duplicated lines among the code dataset provided as input. Therefore, having two snippets of code that are identical but written on a different number of lines, may fool the clone detection algorithm. For this reason, we formatted all methods in our datasets using the IntelliJ formatter~\cite{IntelliJ}, to ensure they all adopt the same coding style (\eg maximum number of characters per line).

We run \textsc{Simian} to identify both types of clones between the code generated by the four models and their respective training set. Then, we compute the percentage of code clones generated by each model by looking at different characteristics of the clones and of the predictions. In particular, we consider in our analysis:

\begin{itemize}

\item \emph{The amount of near-duplicates in the training set}. As previously explained, this has been achieved by training four T5 models on the four datasets we built. We report the percentage of clones generated by each of these models.

\item \emph{The length of the generated code}. We split the generated code blocks in different buckets based on the number of statements they feature (from $\geq2$ to $\geq5$ at steps of 1). Then, we compute the percentage of clones in each bucket. We expect smaller predictions to contain a higher percentage of clones.

\item \emph{The clone type}. When reporting the percentage of clones, we distinguish between Type-1 and Type-2 clones. Type-2 clones will be, by construction, more prevalent, since they are a superset of Type-1 clones, being exact copies but for the different use of identifiers and types.

\item \emph{The training dataset from which the prediction has been ``cloned''}. Each T5 model exploits two training datasets, the pre-training and fine-tuning. We analyze from where the model is more likely to ``copy''.

\item \emph{The correctness of the prediction}. We look whether the models are more likely to ``copy'' when generating correct or wrong predictions. 

\end{itemize}

We complete our study with a correlation analysis in which we assess whether the model is more prone to clone code in specific circumstances. In particular, we use the Spearman test \cite{zar1972significance} to correlate the presence of clones with (i) the \emph{Cyclomatic Complexity} of the test method (ii) the number of lines of the method, and (iii) the \emph{confidence} of the prediction. The first two metrics are computed using the Python Lizard library \cite{lizard}. The last one is a \emph{score} we computed for each prediction made by T5. Such a score $x$ ranges from minus infinity to 0 and it is the log-likelihood of the prediction itself and can be normalized by computing the $exp(ln(x))$. This brings the confidence score between 0.0 and 1.0, with 1.0 indicating the scenario in which the model is confident about the generated prediction.

Finally, it is worth mentioning that, despite the use of the \emph{SentencePiece} tokenizer, the T5 may generate predictions including \emph{unknown tokens}. We performed all the above described analysis both when considering all the instances of the test set as well as when removing the ones that contain at least one \emph{unknown tokens}. The latter are less likely to result in code clones, since the training sets do not contain unknown tokens. We report in the paper the results achieved when excluding from the test set the 6,986 instances for which at least one of the four models generated an unknown token; the results on the whole test set are available in our replication package \cite{replication}, but we anticipate that they are inline with those discussed in the paper. 

\subsection{Replication Package}
The datasets, models, training/testing code, and the achieved raw data are publicly available in our replication package~\cite{replication}.
\section{Results Discussion} \label{sec:results}

To answer our research question, we analyze the percentage of code predictions being \emph{clones} of instances present in each training dataset (\ie pre-training and fine-tuning).

\begin{figure*}[th]
	\centering
	\includegraphics[width=0.9\linewidth]{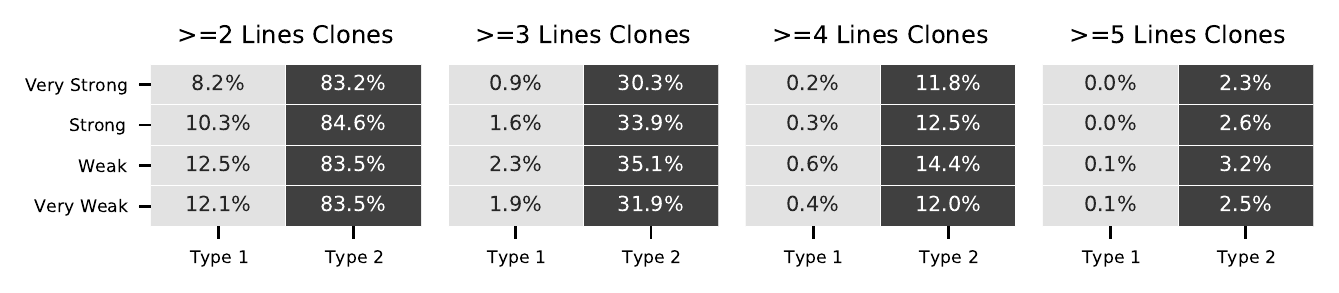}
	\caption{Percentage of Type-1 and Type-2 clones in predictions having different length.}
	\label{fig:results_clones}
\end{figure*}

\smallskip
\textbf{Impact of the amount of near-duplicates in the training set.} \figref{fig:results_clones} shows the percentage of Type-1 and Type-2 clones found in the predictions of the four models fine-tuned on the different datasets (\ie \emph{dataset$_{very\_strong}$}, \emph{dataset$_{strong}$}, \etc). We also organize the predictions in different buckets based on the length of the predicted code (from $\geq$ 2 up to $\geq$ 5). As a reminder, \emph{dataset$_{very\_strong}$} is the smallest (273k instances) and has been built using very strict criteria for the exclusion of near duplicates;  \emph{dataset$_{very\_weak}$} is instead the largest (838k instances), with a higher presence of near duplicates. Note that the gap in dataset size is substantial when moving from the \emph{Very Strong} to the \emph{Strong} configuration (+77\% of instances) and from \emph{Strong} to \emph{Weak} (+47\%), but limited when going from \emph{Weak} to \emph{Very Weak} (+18\%). 

\eject

Before commenting on the achieved results is also worthwhile to remind that the number of generated predictions is 15,742-6,986 for all datasets, since in this results discussion we removed all cases for which at least one of the four models generated an unknown tokens. Thus, a 10\% of clones indicates $\sim$900 generated clones.

With the increase of near duplicates in the training set, the percentage of Type-1 clones in the generated predictions increases as well. This trend is visible across the four datasets with the exception of \emph{dataset$_{very\_weak}$} for which the percentage of generated Type-1 clones is inline with those of \emph{dataset$_{weak}$}. As previously said, this is likely to be due to the small differences in size (\ie in the presence of near-clones) between the two datasets. 

Another trend is also clear when looking at \figref{fig:results_clones}: longer predictions are less likely to be the result of cloned snippets from the training set. If we focus, for example, on the dataset resulting overall in the highest percentage of generated Type-1 clones (\emph{dataset$_{weak}$}), such a percentage moves from 12.5\% for predictions featuring at least two statements, to 2.3\% for those having at least three statements, down to 0.1\% when only considering the generation of non-trivial code snippets composed by at least five statements. Such a result is quite expected since longer predictions are statistically less likely to be equal to instances present in the training set as compared to shorter predictions. 

The analysis of Type-2 clones follows similar trends with a couple of notable differences. First, as expected, the percentage of Type-2 clones is much higher as compared to that of Type-1 clones (from 10 to 100 times higher). This can be explained by the weaker requirements adopted to consider two code snippets as Type-2 clones. Indeed, while the code structure should be similar, two instances are considered as Type-2 clones even if they adopt completely different identifiers and types. Still, confirming what observed for Type-1 clones, the percentage of clones in the predictions steadily drops with the increase in size of the predictions, moving from $\sim$80\% to $\sim$2.5\% when the prediction's size increases from at least 2 to at least 5 lines. Finally, differently from what observed for Type-1 clones, there is no substantial difference in the percentage of Type-2 clones across the different datasets. This may be due to the fact that, differently from Type-1 clones that are verbatim copies of code from the training set, in the case of Type-2 clones the amount of near-duplicates (\ie frequent instances in the training set) is less likely to play a role.

\begin{center}	
	\begin{rqbox}
		\textbf{Take away 1.} Around 10\% of the generated predictions represent a Type-1 clone of code present in the training data, with this percentage going up to $\sim$80\% when considering Type-2 clones. However, the percentage of clones steadily drops when the model is asked to predict several statements, with basically no clones generated when the prediction comprises at least four lines of code. This indicates that advanced tools such as GitHub Copilot that can generate entire functions are very unlikely to ``copy code'' from the training data.
	\end{rqbox}	 
\end{center}


%

\begin{figure}[h!]
	\centering
	\includegraphics[width=\linewidth]{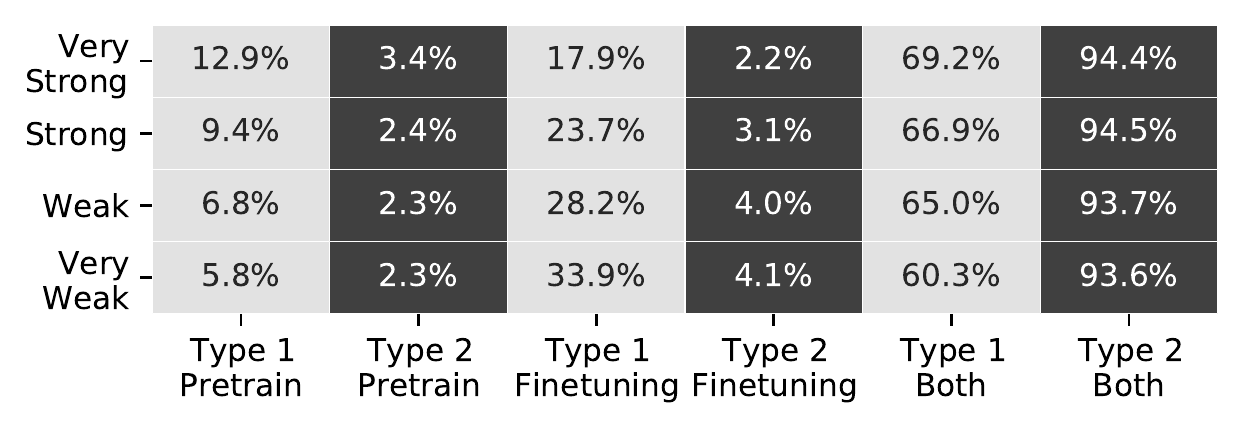}
	\caption{Percentage of Type-1 and Type-2 clones coming from the pre-training and fine-tuning dataset}
	\label{fig:results_pt_ft}
\end{figure}

\smallskip
\textbf{Pre-training and fine-tuning datasets.}
We also analyzed the provenance of the clones, meaning whether the cloned predictions tend to come from the pre-training or from the fine-tuning dataset. \figref{fig:results_pt_ft} summarizes the achieved results. For both types of clones (\ie Type-1 and Type-2) \figref{fig:results_pt_ft} reports the the percentage of clones coming (i) only from the pre-training (\ie the prediction has a clone only in the pre-training dataset); (ii) only from the fine-tuning; and (iii) from both training datasets (\ie both training datasets have at least one instance representing a clone of the prediction). In interpreting the reported percentages it is important to consider that the chances of cloning from the pre-training are expected to be higher, since it contains more ``material'' from which the model can copy. For example, in the \emph{Very Strong} configuration, we have 1.85M methods in the pre-training dataset and 273k methods in the fine-tuning dataset. Therefore, 87\% of the training methods belong to the pre-training dataset and only 13\% to the fine-tuning dataset. The \emph{Very Weak} configuration is the most balanced, still having, however, 69\% of training instances coming from the pre-training dataset against the 31\% of the fine-tuning dataset.

For both types of clones, most of the cloned predictions have a clone both in the pre-training and in the fine-tuning datasets. 

As expected, this is especially true for Type-2 clones, for which almost the totality of cloned predictions follow this pattern. More interesting are the results for Type-1 clones, namely exact copies. While also here there is a substantial percentage of predictions having clones both in the pre-training and in the fine-tuning datasets ($\sim$60\%), most of the predictions only cloned from one of the two training datasets come from the fine-tuning. This result is not really expected considering that the pre-training dataset provides most of the ``training material''. However, possible explanations for this result lie is the order in which the two training phases are applied to the DL model and in their objective. The pre-training changes the weights of the neural network from a random distribution to a distribution able to deal with the pre-training denosing task (\ie guessing masked tokens in the input instance). The tokens masked in the pre-training are randomly selected and may not be contiguous. This means that in very few cases the model sees instances in which it is required to generate complete code statements. The fine-tuning, instead, starts from the pre-trained model or, in other words, from the distribution of weights obtained after pre-training. These weights are then modified to support the fine-tuning task that, in our case, explicitly requires the model to generate complete code blocks. Thus, when the model is asked during testing to generate code blocks, the weights learned during the fine-tuning may come more ``handy'', pushing the model to reuse more knowledge acquired during the fine-tuning rather than during the pre-training.

\begin{center}	
	\begin{rqbox}
		\textbf{Take away 2.} Most of the cloned predictions have clones in both the pre-training and the fine-tuning. For example, in the case of Type-1 clones, $\sim$65\% come from both training datasets. This indicates that code instances seen repeatedly during both training phases are, as expected, more likely to influence the generated predictions. The remaining ones (\ie clones that only come from one of the two training sets), are more likely to come from the fine-tuning. This goes against what we expected since the pre-training is substantially larger than the fine-tuning.
	\end{rqbox}	 
\end{center}

\begin{figure}[h!]
	\centering
	\includegraphics[width=0.8\linewidth]{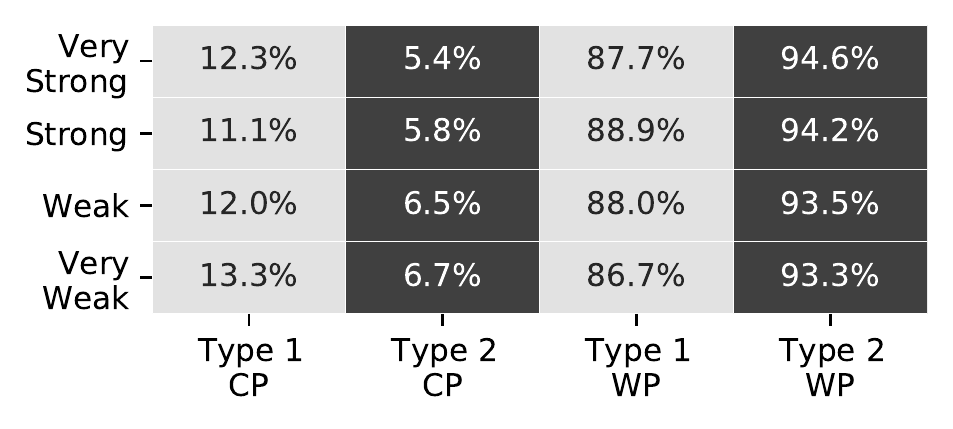}
	\caption{Percentage of Type-1 and Type-2 clones in correct (CP) and wrong (WP) predictions.}
	\label{fig:results_pp_wp}
\end{figure}

\smallskip
\textbf{Correct and wrong predictions.}
\figref{fig:results_pp_wp} shows the percentage of Type-1 and Type-2 clones created by the model when generating correct (CP) and wrong (WP) predictions. Also in this case the percentages must be read by keeping in mind that, across the test instances on which the four models have been tested, all models achieved $\sim$3\% of correct predictions ($\sim$500 cases), implying that we should expect $\sim$3\% of clones being related to correct predictions. Before commenting the results, a few notes on the low percentage of correct predictions achieved by the models (more in our Validity Evaluation section --- \secref{sec:threats}): (i) such a percentage is inline with what has been reported in the literature for applications of DL models to automate other code-related tasks, such as bug-fixing \cite{Tufano:tosem2019} or code review automation \cite{Tufano:icse2021}; (ii) the code generation tasks on which we tested T5 included the generation of code blocks up to six lines of code, thus being non-trivial; (iii) we consider as correct only predictions that are identical to the masked code, meaning that even predictions that are slightly different but semantically equivalent are considered wrong; (iv) finally, as we will discuss in \secref{sec:threats}, our inspection of wrong predictions confirmed that the model learned how to generate syntactically correct code, confirming the validity of the performed training.

The cases in which the predictions represent verbatim copies of code in the training data (Type-1 clones) occur $\sim$4$\times$ more than expected in correct predictions (see \figref{fig:results_pp_wp}), with $\sim$12\% of clones belonging to correct predictions, only representing $\sim$3\% of all predictions. This is likely due to instances (methods) in the training set that, while being different from all methods in the test set, share with them the part of code that we masked. Such a result also rises a warning for the evaluation of DL models in the context of code generation: It might be not enough to just remove duplicates and/or near-duplicates across instances, since the parts of code the model is required to generate in the test set may still be present as verbatim copy in the training set, even when very strict criteria are used to remove duplicates (as in the case of our \emph{Very Strong} configuration). 

Similar observations can be made for Type-2 clones, even though here the percentage of clones in correct predictions is only $\sim$2$\times$ higher than what we expected.

\begin{center}	
	\begin{rqbox}
		\textbf{Take away 3.} Correct predictions are more likely to contain clones of code present in the training datasets. Such a result stresses once more the importance of a proper dataset cleaning to avoid overlaps between the training and the test datasets.
	\end{rqbox}	 
\end{center}

\smallskip
\textbf{Correlation analysis.}
We now discuss the results related to the correlation analysis aimed at understanding if specific characteristics of (i) the code snippet to predict, (ii) the instance containing it, or (iii) the confidence of the model in generating the prediction, may affect the presence/absence of code clones. 

More in details, as explained in \secref{sec:design}, we consider (i) the Cyclomatic Complexity (CC) of the entire instance (method), (ii) the CC of the model's input (\ie the entire method without the block to predict), (iii) the number of lines of the method, (iv) the number of lines of the block to predict, and (v) the confidence of the T5 model. 

We found no significant correlation between all these metrics and the presence of Type-1 clones in the generated predictions. While we identified some significant correlations for Type-2 clones, they are all lower than $|0.15|$, thus indicating very low correlations (complete results in our replication package \cite{replication}).

\smallskip
\textbf{Qualitative Examples.} 
Finally, we present in this section some qualitative examples of the clones generated by the T5 model. Remember that the clones refer only to the part of code predicted by the model, ignoring the surrounding context. In other words, two different methods containing the same block of code predicted by T5 are considered as clones.

\begin{center}	
\begin{lstlisting}[language=Java,caption={Type-1 Clone Generated in Wrong Prediction},captionpos=b]
/* cloned code */
errorFragment = new ErrorFragment();
Bundle msg = new Bundle();
msg.putString("msg", ex.getMessage());
errorFragment.setArguments(msg);
getSupportFragmentManager().beginTransaction()
  .replace(R.id.message, errorFragment).commit();

/* expected prediction */
{ 
errorFragment = new ErrorFragment(); 
Bundle msg = new Bundle(); 
msg.putString("msg", "No or poor internet connection."); 
errorFragment.setArguments(msg); 
getSupportFragmentManager().beginTransaction()
  .replace(R.id.message, errorFragment).commit(); 
}
\end{lstlisting}	
\end{center}

The top part of Listing 1 shows an example of Type-1 clone generated by the T5, while the bottom part reports the correct prediction the model was expected to generate (\ie the code we masked). As it can be seen, the prediction task we asked the model to perform was far from trivial and the T5 managed to generate a prediction really close to the expected one. Such a prediction was the result of cloning four of the five statements to predict from the training dataset. This is one of those cases in which, despite the model did not generate a correct prediction, it went quite close to the target, with the only difference being one of the arguments used in the \texttt{putString} invocation. 

\begin{center}	
\begin{lstlisting}[language=Java,caption={Type-1 Clone Generated in Wrong Prediction},captionpos=b]
/* cloned code */
  if (s != null) 
  {
    try {
      return Integer.parseInt(s);
    } catch (NumberFormatException e) {
  }
}
return null;

/* expected prediction */
{
  Integer interval = OptionAdvanceParser.DEFAULT_INTERVAL; 
  if (optionSet.hasArgument("interval")) { 
    interval = (Integer) (optionSet.valueOf("interval")); 
    if (interval < 1) { 
      throw new IllegalArgumentException("Interval cannot be set below 1.0"); 
    }
  } 
  return interval; 
}
\end{lstlisting}	
\end{center}

Also Listing 2 shows a Type-1 clone generated in a wrong prediction. In this case, we want to put the focus on the fact that the model cloned a snippet of code that represents a quite popular template in Java when converting a \texttt{String} to an \texttt{Integer}. The model understood the need for implementing a check on the value of an \texttt{Integer} but failed in implementing the check that was actually needed (the one shown in the expected prediction). 

\begin{center}

\begin{minipage}{0.95\linewidth}\

\begin{lstlisting}[language=Java,caption={Type-2 Clone Generated in Wrong Prediction},captionpos=b]
/* cloned code */
do {
  CategorySource categorySource = CategorySource.categoryFromCursor(cursor);
  catsSrcs.add(categorySource);
} while (cursor.moveToNext());

/* expected prediction */
{ 
  do { 
  CategorySource categorySource = CategorySource.categoryFromCursor(cursor); 
  catsSrcs.add(categorySource); 
  } while (cursor.moveToNext()); 
}


/* code from the training set */
do {
  CategorySource category = CategorySource.categoryFromCursor(cursor);
  categories.add(category);
} while (cursor.moveToNext());
\end{lstlisting}	
\end{minipage}
\end{center}

Finally, Listing 3 shows an example of Type-2 clones, reporting the prediction on top, the expected code in the middle, and the cloned instance from the training in the bottom. In this case, the model reused the code structure just changing the identifier name from \texttt{category} to \texttt{categorySource}. The model generated something very close to the expected solution, fully relying, however, on code already seen during the training that led it to a wrong prediction.

\section{Validity Discussion} \label{sec:threats}
 
Threats to \emph{construct validity} concern the relationship between the theory and what we observe. In our analysis, similarly with what done in the literature (\eg \cite{HellendoornPGB19,ciniselli2021empirical}), we simulate the usage of the code recommender by masking blocks of code. We are aware that the masked blocks may not completely reflect the way in which developers would use such recommenders in practice.

Another threat is related at the limited prediction performance achieved by our models ($\sim$3\% of correct predictions, detailed results in \cite{replication}). As already discussed, this level of performance is not unusual for the automation of challenging code-related tasks \cite{Tufano:tosem2019,Tufano:icse2021}, such as the code generation task subject of our study. However, a reader may wonder whether the low percentage of clones we observed in the predictions may be due to the fact that, in most of cases, the model just generates garbage, thus not being ``able'' to generate clones of training instances. This is not unusual to observe for DL models trained, for example, on very little data and that tend to generate long meaningless sequences of frequent tokens observed in the training set (\eg sequences of parentheses when dealing with code-related datasets). While we trained our models on millions of Java methods, we inspected a sample of 200 wrong predictions (50 for each of the four models) to manually check whether the generated predictions, while wrong, followed a correct Java syntax. Such an analysis has been performed by the first author, and resulted in only 7 (3.5\%) instances classified ``meaningless predictions'' in terms of Java syntax. 


\eject

Threats to \emph{internal validity} concern factors, internal to our study, that could affect our results. 
The hyperparameters tuning phase described in details in \secref{sec:design} can have a strong impact on the performance of DL models and, consequently, on their likelihood of generating clones. Due to feasibility reasons, we calibrated the hyperparameters only for the \emph{dataset$_{very\_weak}$}, assuming that the chosen configuration would also work well for the other datasets. Hence, it is possible that evaluating a plethora of different configurations may improve the performance of the model and/or impact our findings about the generated clones.

Note also that, as explained, we observed that for some of the instances in the test set the models generated \emph{unknown tokens} that are less likely to result in clones from the training set. For this reason, we reported in the paper the results achieved when ignoring these instances, while the results on the whole dataset are available in our replication package~\cite{replication}. As previously said, the main observed findings still hold when also considering the predictions featuring unknown tokens.

Finally, it is worth mentioning that the accuracy of the \textsc{Simian} clone detector may influence the achieved results. To partially address this threat, the first author inspected 100 randomly selected predictions from our dataset for which Simian identified Type-1 (50 predictions) and Type-2 (50) clones. Since for a single prediction there might be dozens of clones, we inspected 3 clones per each prediction. Concerning Type-1 clones, without surprise, the accuracy was 100\%, since those are predictions being exact copies of code in the training sets. As for Type-2 clones, judging the identified instances was not always straightforward due to the limited size of the identified clones (between 2 and 6 lines). These clones are exact copies at the AST-level and, looking at them from this perspective, we confirm the correctness of all inspected cases. However, especially when looking at very small clones (\eg 2 lines), we acknowledge that some pieces of code, while sharing the same AST structure, may look different to a human, since using different identifiers and/or types.

Threats to  \emph{external validity} are related to the possibility to generalize our results. We chose the T5 model, that showed a strong ability to adapt to different code-related tasks \cite{Mastropaolo:icse2021} achieving remarkable performance. The datasets used in our experiments are all Java-related. We do not claim any generalizability of our findings in terms of adopted DL model and subject programming language. 

We relied on the \textsc{Simian} clone detector \cite{SimScan} that, as explained in \secref{sec:design}, is robust in the processing of syntactically incorrect code. Results may change by using other clone detectors. Moreover, in our study we only considered code clones composed by at least two non-trivial lines. Considering shorter clones, while being an option, is likely to artificially inflate the percentage of clones with instances that are not really relevant in terms of possible licensing issues.
 
\section{Related Work} \label{sec:related}

We discuss the literature related to (i) techniques supporting code completion and (ii) empirical studies investigating circumstances under which clones can be introduced on software systems. Concerning code completion, for the sake of brevity we only focus on DL-based techniques. 

\subsection{DL-based Code Recommendation}

The usage of data-driven techniques for code recommendation has its roots in the works exploiting statistical language models, such as $n$-gram models, to recommend developer with the next code token to write~\cite{Hindle:icse2012,Tu:fse2014,Hellendoorn:fse2017}. Hellendoorn and Devanbu \cite{Hellendoorn:fse2017} also showed that cached $n$-gram models aimed at considering specific characteristics of code (\eg scoped vocabulary) can beat DL-based approaches, and that the two families of techniques can be even combined for better effectiveness.

Later on, Karampatsis \etal~\cite{Karampatsis:DLareBest} suggested instead that neural network models are the best algorithm for code completion. The authors exploited \textit{Byte Pair Encoding}~\cite{bpe} as a strategy to overcome the \emph{out-of-vocabulary problem}, showing that their best model outperforms $n$-gram models and easily adapts to different domains. 

Kim \etal~\cite{kim2020code} leveraged the Transformers neural network architecture for code completion. Using the syntactic information provided by the Abstract Syntax Tree (AST), they were able to fortify the self-attention mechanism. Among the several models they experiment with, the best one reached a Mean Reciprocal Rank (MRR) of 74.1\% in predicting the next token.

Alon \etal~\cite{alon2019structural} proposed a language-agnostic approach named Structural Language Model to tackle the problem of code completion. Based on LSTMs and Transformers, the model leverages the code syntax to represent a given snippet as a tree. They trained their model by providing as input an AST representing a partial expression (statement) missing some consecutive tokens. Their best model reached an accuracy of 18\% for top predictions.

Svyatkovskiy \etal~\cite{svyatkovskiy2020intellicode} introduced IntelliCode Compose, a multilingual code completion tool able to predict multiple tokens of arbitrary types. They use subtokens to overcome the \emph{out-of-vocabulary problem} without leveraging high-level structural information like AST. Their model, trained to predict entire statements of Python language, achieves a perplexity of 1.82.

A Transformer-based architecture was presented by Liu \etal~\cite{Liu:ase2020}. They pre-trained their model to incorporate both code understanding and generation tasks. Then, they fine-tuned it on the classic code completion task, predicting the next token the developer is likely to write.

Svyatkovskiy \etal \cite{svyatkovskiy2021fast} tackled code completion from a different perspective, shifting the problem from the generation of code to the ranking of proposed solutions. They used static analysis, a cheaper algorithm in terms of memory footprint than  generative models, to provide a list of recommendations to their DL model, which learns to rerank them.

Jian \etal \cite{li2017code} proposed a pointer mixture deep learning model for Python that leverages the pointer copy mechanism to address the out-of-vocabulary problem. The idea behind the proposed architecture is to pick an out-of-vocabulary word from the local context through a pointer component when it is not possible to generate a within-vocabulary token.

Recently, Aye and Kaiser \cite{aye2020sequence} proposed a model able to predict the next top-$k$ tokens while taking into account (i) prediction latency, (ii) size of the model and its memory footprint, and (iii) validity of suggestions. These are all actual constraints for real-world architectures.  

Chen \etal \cite{chen2020holistic} exploited a DL model for API recommendation, merging structural and textual code information retrieved from the API context graph and the code token network. Their model significantly outperforms the existing statistical and DL approaches for API recommendation based on graphs and trees.


Chen \etal introduced Copilot \cite{chen2021evaluating}, a new GPT model trained on more than 150Gb of data from GitHub. Their trained model achieved state-of-the-art performance in the demanding task of predicting the whole method when providing as input to the model just a natural language description of what the developer wants to implement. Since the dataset used to train Copilot is not publicly available, in our work we adopted another recent DL-based approach proposed by Ciniselli \etal \cite{ciniselli2021empirical}, that has been already previously described.


\subsection{Studies Investigating the Introduction of Code Clones}

A plethora of techniques have been proposed in the literature to identify code clones in software systems (\ie code clone detection). These techniques exploit information extracted from the AST \cite{baxter1998clone,SimScan,bulychev2008duplicate}, use a graph-based representation of code \cite{komondoor2001using,higo2011code,krinke2001identifying}, or look at token-based similarities \cite{kamiya2002ccfinder,CCFinderX,kawaguchi2009shinobi}. A complete coverage of such a topic can be be found in the literature review by Rattan \etal \cite{rattan2013software}. In this section, we focus on studies investigating the introduction of code clones, since those are the most related to our work.

Kim \etal \cite{kim2004ethnographic} investigated the reasons behind the copy and paste mechanism used by developers. Although in 74\% of cases the copy involve less than one line of code (\ie the copy of a single variable or expression), the copy of snippets of code is still frequent ($\sim$25\%) and sometimes dictated by the programming language itself, pushing the copy of syntactic templates. Even the difficulty in understanding large systems fosters the copying of functionalities implemented in the past. 

Li \etal \cite{li2006cp} showed that copy-pasting code can reduce the programming effort. Moreover, copying a code snippet instead of abstracting it in a new function can reduce overhead in execution, even though sometimes this may introduce some bugs, generally due to errors during the copy. Also Jiang \etal \cite{jiang2007context} highlight as developers feel less comfortable with completely new code that may introduce subtle bugs, and prefer code reuse when possible.


Kapser and Godfrey \cite{kapser2008cloning} showed that developers significantly use code clones while developing software, and highlighted good motivations behind such a practice. For example, code clones are sometimes introduced as side effect of programmers' memory, since developers tend to repeat common solutions without knowing that they are already present in the source code. This can also improve the readability of source code. Moreover, when the same component interacts with lots of different systems, forking the code (\ie copying a part of code that evolves independently from the remaining code base) is a safe solution that can avoid refactoring costs and bug introduction.

The seek of simplicity has also been reported as a motivation for clone introduction by Baker~\cite{baker1995finding}. The author also reported that the process management might encourage duplication since developers' performance is evaluated based on how much new code they write, which does not promote the refactoring of old code.

To the best of our knowledge, the study mostly related to the one we conducted has been presented by Albert Ziegler in a blog post about GitHub Copilot\footnote{\url{https://docs.github.com/en/github/copilot/research-recitation}}. The author investigated the extent to which GitHub Copilot suggestions are copied from the training set they used and concluded that Copilot rarely recommends pieces of code that are verbatim copies of instances from the training set, and when this happens these are coding solutions recurring across several systems, that also humans are likely to reuse. Considering the complexity of the recommendations generated by Copilot (\ie up to entire functions), our results support such a conclusion since we found that, when the DL model is used to recommend several code statements, it is very unlikely to provide cloned code as output.
\section{Conclusion} \label{sec:conclusion}

Code recommenders are becoming more and more popular. GitHub Copilot \cite{chen2021evaluating} substantially pushed the adoption of these tools by developers, making central questions related to possible licensing issues that may come from their adoption in a commercial setting. Indeed, these tools are usually trained on open source code, the usage of which is regulated by FOSS licenses. 

While the posed licensing questions are part of a non-trivial debate among the open source community, researchers, and tool builders, we started investigating a concrete and related research question: \emph{To what extent do DL-based code recommender systems generate predictions by cloning code snippets from the training set?}

We trained a Text-To-Text-Transfer-Transformer (T5) model on over $\sim$2M Java method with the task of recommending code blocks aimed at finalizing the methods' implementation. Then, we used a clone detector to check whether the predictions it generated on the test set represent Type-1 or Type-2 clones of instances in the training datasets. Our findings show that, while for short predictions the trained deep learning model is likely to generate a non-trivial percentage of clones ($\sim$10\% Type-1 and $\sim$80\% Type-2), such a percentage quickly approaches 0\% when the model generates more complex predictions composed my at least four code statements.

In summary, our findings suggest that the likelihood of obtaining clones generated by DL-based code recommenders that are possibly ``harmful'' in terms of licensing issues is extremely low.

Our future work will mainly focus on replicating our work on larger, more performant, DL-based code recommenders and different clone detectors, with the goal of confirming or confuting our findings.

The material used in this study is publicly available \cite{replication}.

\smallskip

\section*{Acknowledgments}
This project has received funding from the European Research Council (ERC) under the European Union's Horizon 2020 research and innovation programme (grant agreement No. 851720).

\eject

\bibliographystyle{ACM-Reference-Format}
\bibliography{main}

\end{document}